\documentclass[twocolumn,showpacs,preprintnumbers,amsmath,amssymb]{revtex4}

\usepackage{graphicx}             

\begin{document}


\title{Fermi surface nesting and charge-density wave formation in rare-earth tritellurides}

\author{J.~Laverock, S.~B.~Dugdale, Zs.~Major, and M.~A.~Alam}

\affiliation{H.~H.~Wills Physics Laboratory, University of Bristol, Tyndall 
Avenue, Bristol BS8 1TL, United Kingdom}

\author{N.~Ru and I.~R.~Fisher}

\affiliation{Geballe Laboratory for Advanced Materials and Department of
Applied Physics, Stanford University, Stanford, California 94305-4045}

\author{G.~Santi}

\affiliation{Department of Physics and Astronomy, University of Aarhus,
DK-8000 Aarhus C, Denmark}

\author{E.~Bruno}

\affiliation{Department of Physics, University of Messina, Salita Sperone 31,
98166 Messina, Italy}

\date{\today}

\begin{abstract}
The Fermi surface of rare-earth tri-tellurides ({\it R}Te$_{3}$) is investigated
in terms of the nesting driven
charge-density wave formation using positron annihilation and first-principles
LMTO calculations.
Fermi surface nesting is revealed as a strong candidate for driving charge-density
wave formation in these compounds. The nesting vector obtained from positron annihilation
experiments on GdTe$_{3}$ is determined to be ${\mathbf q}~=~(0.28\pm0.02,0,0)~{\mathbf a}^{*}$,
(${\mathbf a}^{*}=2\pi/{\mathbf a}$), in excellent agreement
with previous experimental and theoretical studies.
\end{abstract}


\maketitle

\section{Introduction}
The existence of charge-density wave (CDW) ground states in low
dimensional metals has attracted a great deal of interest over many 
years (see for example \cite{wilson:74,friend:79,carpinelli:96}).
It is well known that instabilities in the Fermi surface
(FS) due to particular features of the electronic band structure (e.g. nesting
features, van Hove saddle points) can lead to the 
emergence of new ground states such
as spin-density \cite{overhauser:62} or charge-density waves
\cite{wilson:86}. Examples of systems where this is thought to be
the case include
the transition metal dichalcogenides \cite{wilson:74}, pure Cr
\cite{overhauser:62} and organic charge-transfer salts \cite{friend:79}.
While the underlying electronic band structure of a system whose new periodicity is
commensurate with the lattice is well understood, the behavior in cases
where the periodicities are incommensurate is much less clear \cite{voit:01}. 

Motivated by recent angle-resolved photoemission (ARPES) studies aimed at
clarifying this issue \cite{brouet:04,komoda:04}, we have combined first-principles 
electronic structure calculations and positron annihilation experiments
to establish the FS topologies of a model CDW system, namely the
rare earth tritellurides.

In systems whose FS topologies include large parallel faces spanned by
a (nesting) vector ${\mathbf q}$, there will be an instability in that system towards
opening up gaps on these faces by introducing some new periodicity and hence new superzone
boundaries. A CDW is an example of the results of such an instability.
Although in general the CDW
will not be commensurate with the lattice (since its periodicity
is dictated by the nesting vector), many CDWs experience ``lock-in''
to a suitably commensurate period, thereby lowering the strain
energies associated with the distortion.

\begin{figure}[p]
\begin{center}
\includegraphics[width=0.75\linewidth,clip]{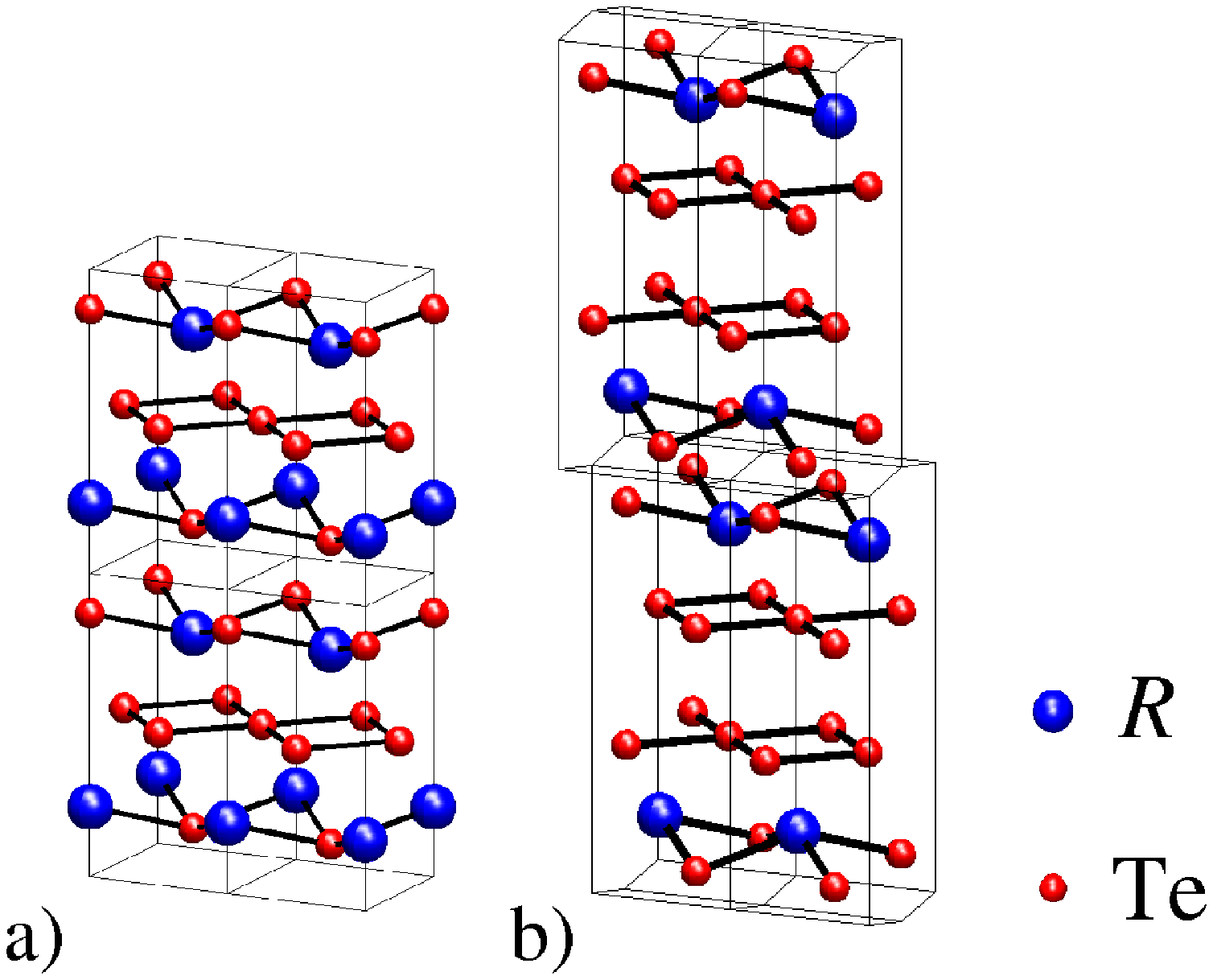}
\end{center}
\caption{\label{struc} (color online) The structure of (a) {\it R}Te$_{2}$ and
(b) {\it R}Te$_{3}$. Note the two Te planes sandwiched between $R$Te layers
in $R$Te$_{3}$.}
\end{figure}

The rare-earth tritelluride series, {\it R}Te$_{3}$ ({\it R} = Y, La -- Tm),
crystallises in the layered NdTe$_{3}$ structure \cite{norling:66},
and consists of sheets of
inert {\it R}Te slabs sandwiched between a double-layer of square
planar Te sheets (see Fig.\ \ref{struc}). It should be noted that
the crystollagraphic {\it b}-axis is the long axis. 
Resistivity measurements have revealed a large anisotropy
($\rho _{a} / \rho _{b}$) across the series, as much as
$\sim$~5000 in the case of SmTe$_{3}$ \cite{dimasi:94}.
The electron diffraction results of DiMasi {\it et al.} showed
strong satellite peaks, which were interpreted as stemming from the periodic
lattice distortions associated with the presence of 
an incommensurate CDW \cite{dimasi:95}. They report
a range of wavevectors ($q \approx 0.27 - 0.31~a^{*}$, $a^{*}=2\pi/a$) (i.e.
around $(2/7)~a^{*}$) for {\it R}Te$_{3}$,
where the CDW was found to be stable for all the rare-earths, $R$ (from
La to Tm), and
occurred for most materials along both the ${\mathbf a}$ and 
${\mathbf c}$ axes. In some samples, CDW formation was only
observed along one of these directions, although this was found to be
highly sample dependent.
ARPES studies \cite{gweon:98,brouet:04} have successfully mapped the
gap anisotropy, finding the FS to be very significantly gapped
($\sim$~200~meV) along $\Gamma - Y$ ($\Gamma - Z$ in their
coordinates); this would correspond to a temperature well beyond the
samples' melting points.
Further,
the ARPES studies have confirmed the magnitude of the CDW
wave-vector to be $q \approx (2/7)~a^{*}$, equal to
that observed by DiMasi {\it et al.} \cite{dimasi:95}.
The {\it R}Te$_{3}$ system thus presents an unprecedented
opportunity to study 
the correlation between the FS topology (through its
nesting) and the CDW period.

The rare-earth ditellurides {\it R}Te$_{2}$ (Fig.\ \ref{struc})
are a related material,
crystallising in the layered tetragonal anti-Cu$_{2}$Sb structure
\cite{wang:66}.
In these materials, electron diffraction reveals satellite peaks
corresponding to a commensurate
superstructure characterised by a $(2 \times 1 \times 1)$ supercell,
attributed to CDW formation \cite{dimasi:96}.
Calculations of the electronic structure of this system will be
presented alongside those for {\it R}Te$_{3}$
in order to highlight the similarities and differences.

Here we present the FS of GdTe$_{3}$ as measured by 2-dimensional
angular correlation of electron-positron annihilation radiation (2D-ACAR)
in conjunction with first-principles electronic structure calculations
of lutetium di- and tri-tellurides.

\section{Electronic structure calculations}
The electronic band structure
was calculated for both lutetium di- and tritelluride
using the linearized muffin-tin orbital (LMTO) method within the atomic sphere
approximation (ASA), including combined-correction terms
\cite{barbiellini:03}. Lu was chosen to avoid the complications
associated with the description of $f$-electrons within the local
density approximation.
The parameters of the crystalline structure 
were 4.55\AA~($b/a = 2.037$)~and~4.34\AA~($b/a = 5.93$)~respectively,
corresponding to those found for LaTe$_{2}$ in Ref. \cite{dimasi:96}
and SmTe$_{3}$ in Ref. \cite{dimasi:94}. The results of the calculation
were found to be relatively insensitive to changes in the lattice
parameter of $\sim 5$\%, allowing us
to address all the different rare-earth systems \cite{structure}. All calculations used  
a basis of $s$, $p$, $d$ and $f$ states, and self-consistency was
achieved at 1568 and 1176 k-points respectively in the irreducible Brillouin
zone (BZ).

Although the tritellurides are found to have
very slightly orthorhombic unit cells ($a \neq c$), a square-based cell
($a = c$) is adopted (since the effects were found to be insignificant).
It is worth highlighting that there will still be in-plane anisotropy in the
tritellurides even for
$a = c$, caused by the {\it R}Te layer orientation, which will be reflected
in the electronic structure.

\subsection{Lutetium ditelluride (LuTe$_{2}$)}
\begin{figure}[b]
\begin{center}
\includegraphics[width=1.00\linewidth,clip]{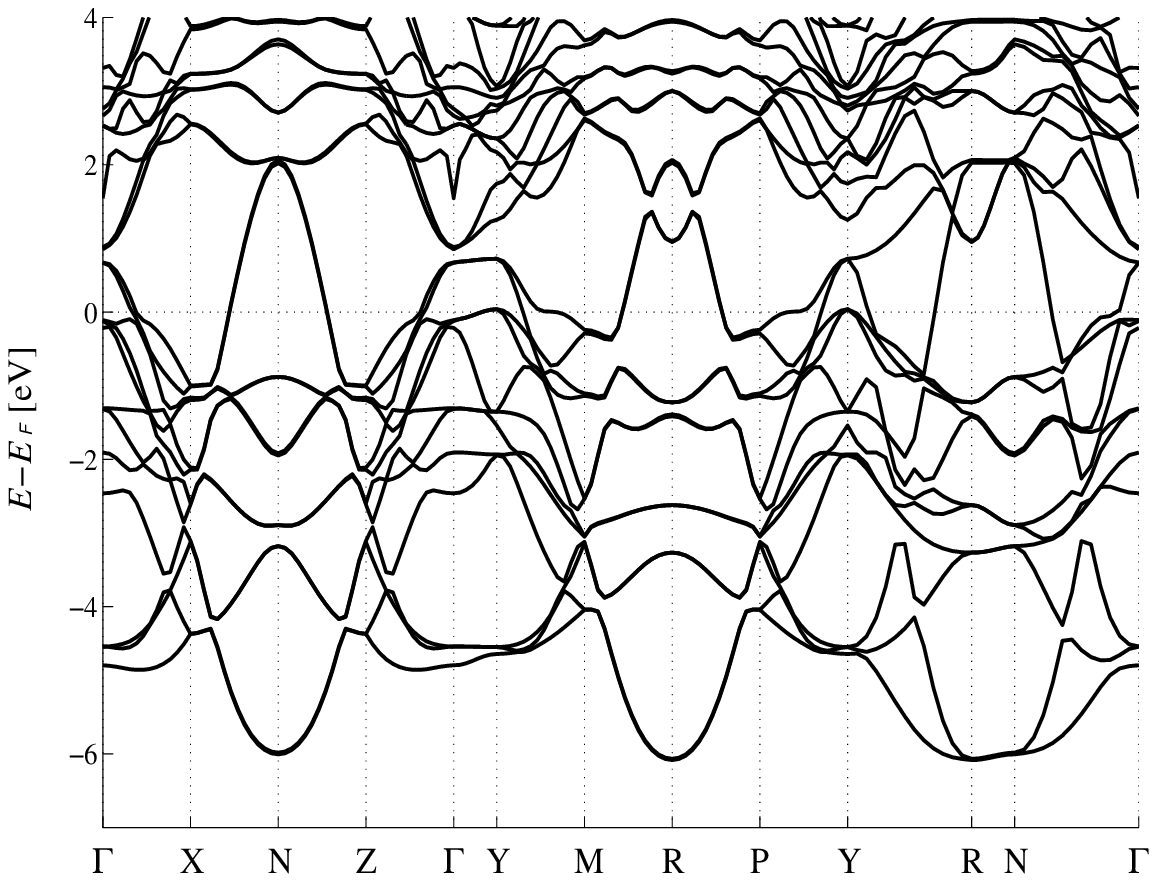}
\end{center}
\caption{\label{lu2bs} The electronic band structure of LuTe$_{2}$.}
\end{figure}

\begin{figure}[b]
\begin{center}
\includegraphics[width=1.00\linewidth,clip]{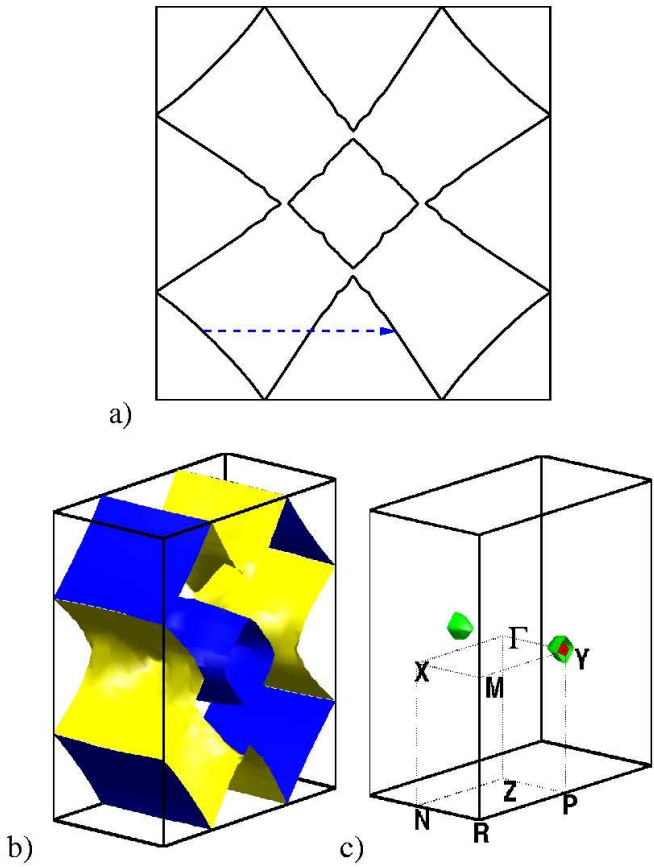}
\end{center}
\caption{\label{lu2fs} (color online) The calculated Fermi surface of
LuTe$_{2}$
(a) on a (010) plane through $\Gamma$ with the proposed nesting vector illustrated
(for clarity the $Y$ pockets have been omitted here), (b) the FS sheets
responsible for the nesting, and (c) the $Y$-centered hole pockets.
High symmetry points have been labeled in (c). The $\Gamma$-point is located
at the center of the zone.}
\end{figure}

The electronic band structure and FS for LuTe$_{2}$ is shown in
Figs.\ \ref{lu2bs} and \ref{lu2fs} respectively. As expected, there is little
electronic interaction between layers in these compounds, expressed
in the small dispersion along [010] (e.g. $\Gamma - Y$ and $N - R$). We
observe three bands to cross the Fermi level ($E_{\mbox{F}}$), principally due to 
$5p$ orbitals of the planar Te layer. Two of these bands result in the
diamond-shaped FS, also predicted by
previous calculations \cite{dimasi:96,kikuchi:97,shim:04}.
These are the sheets which show the nesting.
Two further FS sheets are observed in the present calculation, forming small hole
pockets centered at $Y$ (see Fig.\ \ref{lu2fs}c). As may be seen in
Fig.\ \ref{lu2bs}, the bands responsible for these two sheets only just
cross $E_{\mbox{F}}$, and thus the presence or absence of these sheets
is likely to be very sensitive to the details of the calculation (such
as the 
description of the Lu $4f$ electrons in the local density approximation).
It is emphasised that
these features are irrelevant to the nesting vector previously
proposed by DiMasi {\it et al.}
\cite{dimasi:96}, corresponding to a commensurate nesting vector
of $q~=~(1/2)~a^{*}$, shown in Fig.\ \ref{lu2fs};
our results are consistent with the existence of such a nesting vector.

\subsection{Lutetium tritelluride (LuTe$_{3}$)}
\begin{figure}[b]
\begin{center}
\includegraphics[width=1.00\linewidth,clip]{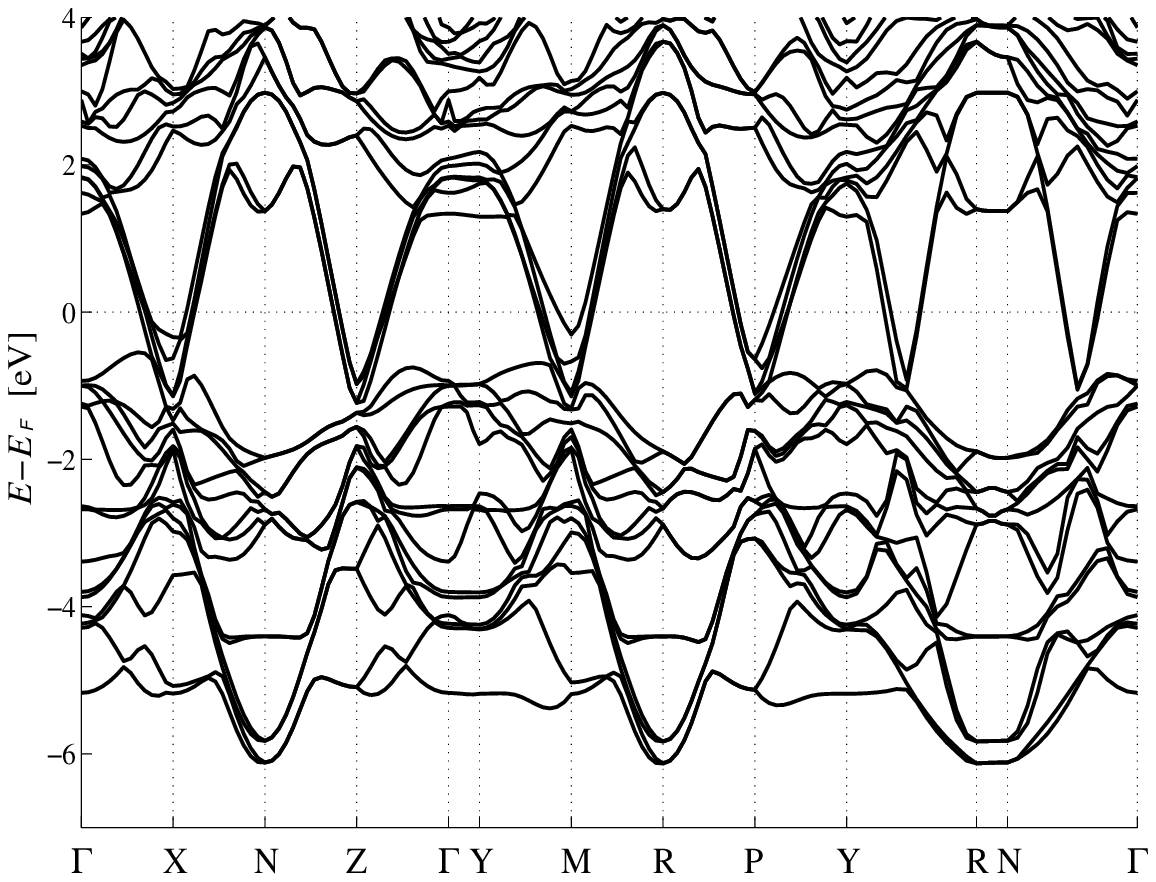}
\end{center}
\caption{\label{lu3bs} The electronic band structure of LuTe$_{3}$.}
\end{figure}

\begin{figure}[b]
\begin{center}
\includegraphics[width=1.00\linewidth,clip]{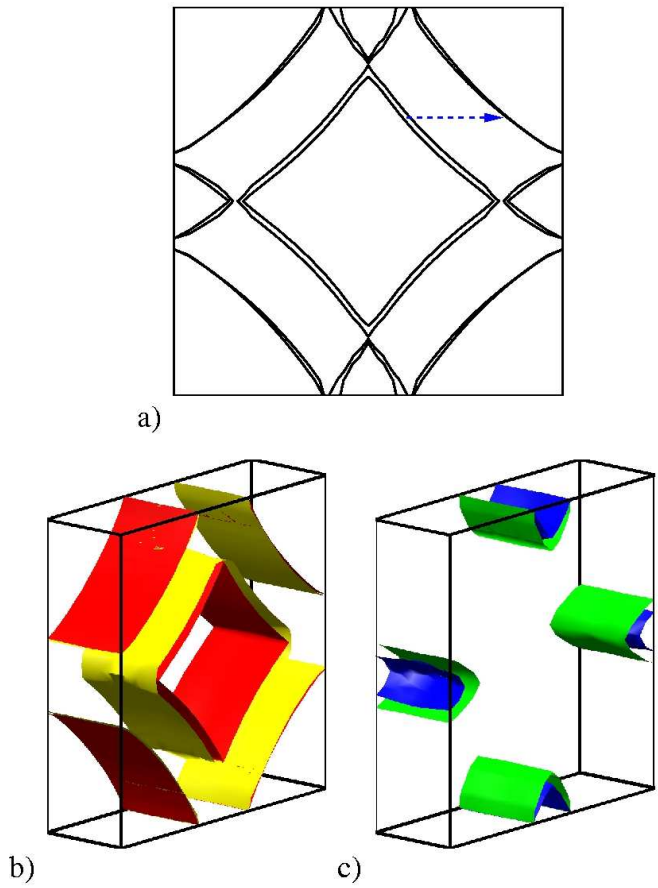}
\end{center}
\caption{\label{lu3fs} (color online) The calculated Fermi surface of
LuTe$_{3}$ (a) on a (010)
plane through $\Gamma$ with the proposed nesting illustrated,
(b) the two diamond shaped sheets (split by the coupling between the 
two Te planes) which exhibit strong nesting, and (c) the two $X$ pockets.
The symmetry points are the same as those labelled in Fig.\ \ref{lu2fs}.}
\end{figure}

The LuTe$_{3}$ band structure (Fig.\ \ref{lu3bs}) differs most notably
from LuTe$_{2}$ in the appearance of a bilayer splitting induced as a
result of the coupling between the two Te planes. This is observed 
as a small splitting of the sheets which compose the FS
(Fig.\ \ref{lu3fs}), and which has been reported experimentally in ARPES
studies for CeTe$_{3}$ and SmTe$_{3}$ \cite{gweon:98,brouet:04}.

\begin{figure}[b]
\begin{center}
\includegraphics[width=1.00\linewidth,clip]{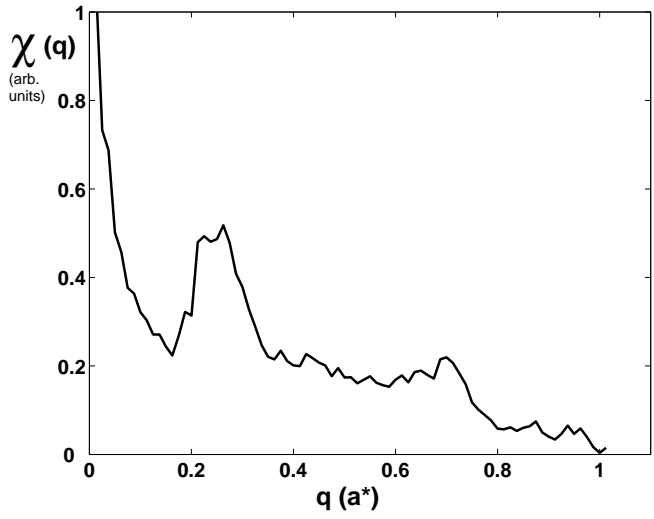}
\end{center}
\caption{\label{chiq} The generalized susceptibility $\chi(q)$ calculated
using the LMTO method for LuTe$_{3}$ along the [100] direction,
showing the broad peak around $q \approx 0.26~a^{*}$ due to nesting between the two diamond shaped sheets.}
\end{figure}

Fig.\ \ref{lu3fs}a shows the nesting vector previously proposed for these
materials, $q \sim (2/7)~a^{*}$, by DiMasi {\it et al.} \cite{dimasi:95},
observed directly in ARPES studies \cite{brouet:04,gweon:98}. It is clear from
this figure that our band structure results strongly support such potential
for nesting, further emphasised by the calculation of $\chi(q)$, for ${\mathbf q}$ along [100]
(see Fig.\ \ref{chiq}). This displays a clear, broad peak centered at
$q \approx 0.26~a^{*}$ (the width of which arises due to the
splitting of the FS), corresponding to nesting between the inner and outer
diamond-shaped features (as shown in Fig.\ \ref{lu3fs}).

Since the FS is composed of states associated with the Te layer,
no appreciable difference is likely to be observed in the nesting features of $R$Te$_{3}$ 
for different rare earth atoms, indicated by the stability of the wavevector
for different $R$ observed
by DiMasi {\it et al.} \cite{dimasi:95}, and so comparisons may be made between
the lutetium compounds used in the calculation and the gadolinium compounds
used in the experiment (and, for that matter, the rare-earths in other studies).

\section{Positron annihilation measurements of the Fermi surface}
The occupied momentum states, and hence the FS, can be accessed via the
momentum distribution using the 2D-ACAR technique
\cite{west:95}.
This well-established technique has recently been used
by some of the present authors to determine the FS topology and identify
nesting features in a wide range of systems
\cite{dugdale:97}.
A 2D-ACAR measurement yields a 2D projection
(integration over one dimension) of the underlying momentum distribution.
Given that the {\it R}Te$_{3}$ series is
strongly two-dimensional, the dispersion along the
$\Gamma - Y$
direction is small and thus the FS is strongly two dimensional; in this
case one single projection along [010] is sufficient to observe the
topology of the FS.
Application of the Lock-Crisp-West (LCW) prescription
\cite{lock:73}, where the ${\mathbf p}$-space momentum density is `folded'
back into the first BZ, yields a translationally
invariant ${\mathbf k}$-space distribution which indicates the
level of occupancy of the states across the (projected) BZ.
The FS topology expresses itself through changes in
this occupancy according to whether a particular ${\mathbf k}$-state is occupied or
unoccupied for each band. The virtue of
the 2D-ACAR technique in such studies is that it reveals {\it directly}
the shape of the FS through the shape of the occupied and unoccupied
regions. The effect of a charge-density wave gap on
the momentum distribution will be to slightly smear the step associated with the
presence of the FS; however, given that the intrinsic momentum resolution
of the spectrometer is $\sim$ 15\% of the BZ, this additional 
smearing is negligible and hence the 2D-ACAR spectrum is insensitive
to the presence of the CDW.
The positron annihilation technique
thus reveals the FS unreconstructed by the CDW (i.e. the ungapped FS).

The positron annihilation measurements were performed using the Bristol
spectrometer. The single
crystal samples were grown by slow cooling of a binary melt containing
an excess of Te, as described previously by Ref. \cite{brouet:04}.
The GdTe$_{3}$ sample was oriented to obtain a projection along the [010]
crystallographic direction, and $\sim$ 150 million counts were accumulated
at a sample temperature of $\sim$ 80K. A maximum entropy based
deconvolution routine was used to reduce the smearing caused by the
finite instrument resolution \cite{dugdale:94}.

\begin{figure}[b]
\begin{center}
\includegraphics[width=1.00\linewidth,clip]{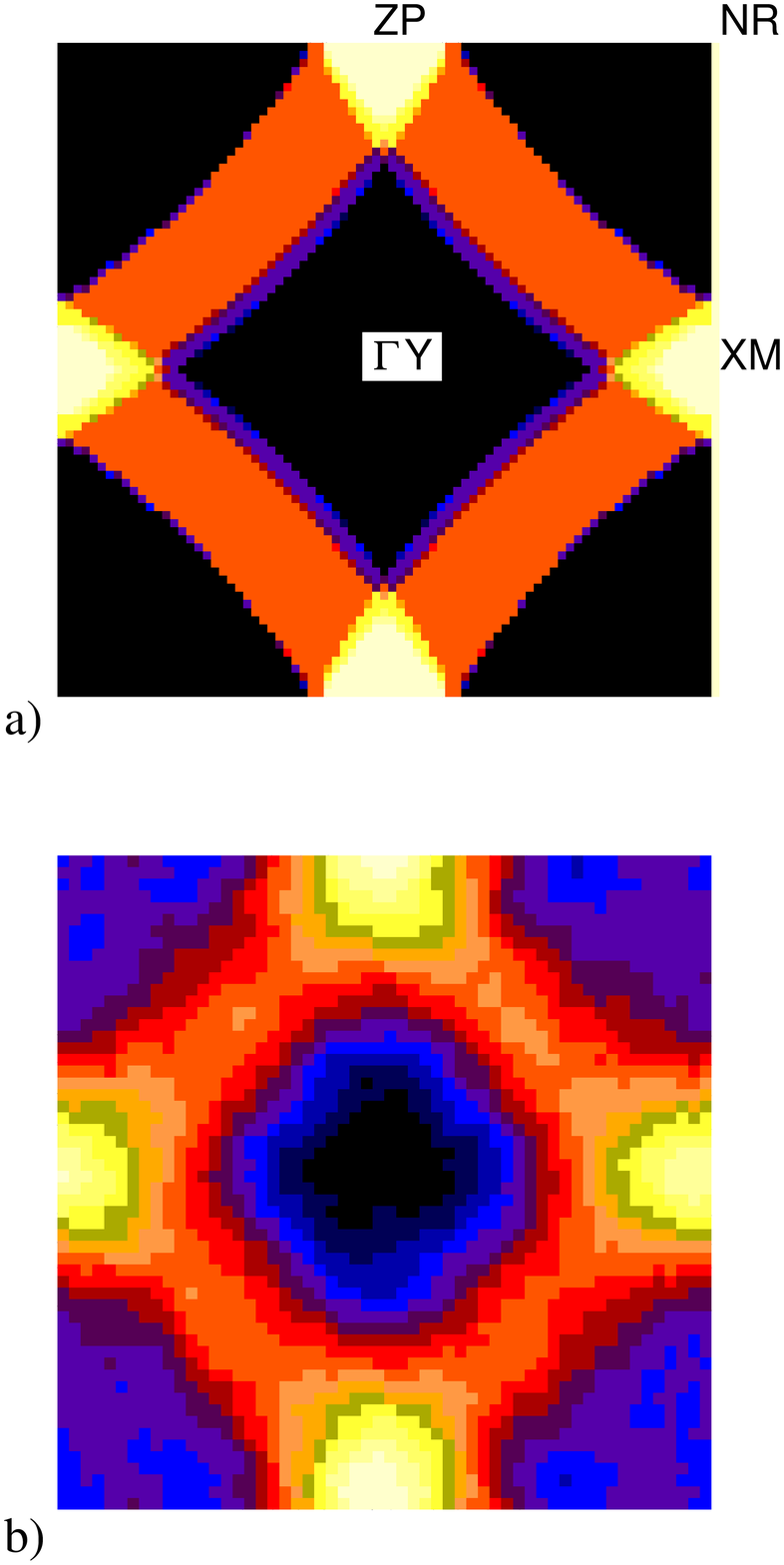}
\end{center}
\caption{\label{emd} (color online)
(a) Projected occupation density obtained from the LMTO calculation
for LuTe$_{3}$, and
(b) occupation density within
the first BZ for GdTe$_{3}$ measured by the 2D-ACAR experiment.
In both cases, lighter colors (shades) represent higher occupancy, and high
symmetry points have been labelled in projection.}
\end{figure}

The ${\mathbf k}$-space occupation density,
projected down the [010] direction,
is presented in Fig.\ \ref{emd}, alongside the occupancy 
obtained from the LMTO calculation.
Further measurements of TbTe$_{3}$ were
also performed at a lower statistical precision,
revealing identical features to GdTe$_{3}$, and will not be discussed
any further.
Excellent agreement is observed between the experimental
and theoretical BZ occupancy; the $X$ electron pockets are clearly
visible in the data, and agree in both size and shape with
the theoretically determined pockets.
Whereas in the ARPES studies of Ref.~\cite{gweon:98}
the FS was observed to be gapped along $\Gamma - Y$
($\Gamma - Z$ in their coordinates), in this study no such gapping
is observed, the effect of the CDW being negligible when compared with
the momentum resolution of the apparatus (as described above).

As the LMTO calculation highlights, the Te bands
that cross the Fermi level are degenerate
for the ditellurides. However, owing to coupling between the two
non-equivalent square planar Te sheets in the tritellurides, these
split. Whilst this splitting has been observed 
(for example, \cite{gweon:98,brouet:04}), it should be pointed
out that our momentum resolution
precludes observation of this small splitting
of the FS in our experiment.
Note, however, that the splitting, being smaller than the experimental resolution, is likely
to enhance the observability of the nested Fermi surface.

The interpretation of the data in terms of nesting provides compelling
evidence in support of the role of the FS in driving CDW formation.
Strong nesting features, identical to those outlined in the
calculated FS, are clearly visible in the data, connecting large
parallel surfaces of FS (Fig.\ \ref{emd}), and revealing a nesting
vector (present in both [100] and [001] directions) connecting
optimally nested portions of the experimental FS of
${\mathbf q}~=~(0.28\pm0.02,0,0)~{\mathbf a}^{*}$.
(The Fermi surface breaks have been determined from the experimental data by a ``zero crossing''
method \cite{dugdale:94}, where the ``raw'' k-space occupancy map has been subtracted from the ``Maximum Entropy
deconvoluted'' k-space occupancy of the raw data. The zero-level contour reveals the breaks
due to the diamond-shaped sheets, from which the nesting vector has been extracted.)
This compares favorably with the superlattice
reflections observed in electron diffraction experiments
\cite{dimasi:95}, and with
nesting vectors of $q \approx (2/7)~a^{*}$
revealed via ARPES studies \cite{gweon:98,brouet:04}.

\section{Conclusion}
In conclusion, we have presented the experimental ungapped Fermi surface of
the high-temperature CDW compound GdTe$_{3}$.
Complementary band structure calculations on LuTe$_{3}$
and the related compound, LuTe$_{2}$, are also presented.
Both the experimental and theoretical
FS topologies display excellent propensity for nesting, supporting
the role of the Fermi surface in driving the formation of the CDW in these
compounds. The nesting features are found to be in excellent agreement
with previously proposed experimental \cite{dimasi:95,gweon:98,brouet:04}
and theoretical \cite{dimasi:95,kikuchi:97,shim:04} studies on the {\it R}Te$_{3}$
series, revealing an experimental nesting vector of~
${\mathbf q}~=~(0.28\pm0.02,0,0)~{\mathbf a}^{*}$.

\section{Acknowledgements}
We acknowledge the financial support of the UK EPSRC and the
Royal Society (S.B.D.). I.R.F. and N.R. ackowledge support by the Department of Energy, Office
of Basic Energy Sciences under contract DE-AC03-76SF00515. I.R.F. is
also supported by the Alfred P. Sloan Foundation.
G.S. ackowledges partial funding from 
the European Commission Human Potential Programme under Contract No. HPRN-CT-2002-00295, `Psi-k 
f-electron'.
We would also like to thank V. Brouet for useful discussions.



\begin{thebibliography}{99}

\bibitem{wilson:74}
J.~A.~Wilson, F.~J.~Di Salvo and S.~Mahajan, Phys. Rev. Lett. {\bf 32}, 882 (1974).

\bibitem{friend:79}
R.~H.~Friend and D.~J\'{e}rome, J. Phys. C: Sol. St. Phys. {\bf 12}, 1441 (1979).

\bibitem{carpinelli:96}
J.~M.~Carpinelli, H.~H.~Weltering, E.~W.~.Plummer and R.~Stumpf, Nature {\bf 381},
398 (1996)

\bibitem{overhauser:62}
A.~W.~Overhauser, Phys. Rev. {\bf 128}, 1437 (1962).

\bibitem{wilson:86}
R.~L.~Withers and J.~A.~Wilson, J. Phys. C: Sol. St. Phys. {\bf 19}, 4809 (1986).

\bibitem{voit:01}
J. Voit, L. Perfetti, F. Zwick, H. Berger, G. Margariondo, G. Gr\"uner, H. H\"ochst and M. Gironi, Science {\bf 290}, 501 (2001).

\bibitem{brouet:04}
V.~Brouet, W.~L.~Yang, X.~J.~Zhou, Z.~Hussain, N.~Ru, K.~Y.~Shin, I.~R.~Fisher
and Z.~X.~Shen, Phys. Rev. Lett. {\bf 93}, 126405 (2004).

\bibitem{komoda:04}
H.~Komoda, T.~Sato, S.~Souma, T.~Takahashi, Y.~Ito and K.~Suzuki, Phys. Rev. B {\bf 70}
195101 (2004).

\bibitem{norling:66}
B.~K.~Norling and H.~Steinfink, Inorg. Chem. {\bf 5}, 1488 (1966).

\bibitem{dimasi:94}
E.~DiMasi, B.~Foran, M.~C.~Aronson and S.~Lee, Chem. Mater. {\bf 6}, 1867 (1994).

\bibitem{dimasi:95}
E.~DiMasi, M.~C.~Aronson, J.~F.~Mansfield, B.~Foran and S.~Lee, Phys. Rev. B.
{\bf 52}, 14516 (1995).

\bibitem{gweon:98}
G.-H.~Gweon, J.~D.~Denlinger, J.~A.~Clack, J.~W.~Allen, C.~G.~Olson, E.D.~DiMasi,
M.~C.~Aronson, B.~Foran and S.~Lee, Phys. Rev. Lett. {\bf 81}, 886 (1998).

\bibitem{wang:66}
R.~Wang, H.~Steinfink and W.~F.~Bradley, Inorg. Chem. {\bf 5}, 412 (1966).

\bibitem{dimasi:96}
E.~DiMasi, B.~Foran, M.~C.~Aronson and S.~Lee, Phys. Rev. B. {\bf 54}, 13587
(1996).

\bibitem{barbiellini:03}
B. Barbiellini, S.~B. Dugdale and T. Jarlborg, Comp. Mater. Sci. {\bf 28},
287 (2003).

\bibitem{shim:04}
LSDA+U calculations have recently been used to avoid $f$ electrons issues
within the LDA. See 
J.~H.~Shim, J.-S.~Kang and B.~I.~Min, Phys. Rev. Lett. {\bf 93} 156406 (2004).

\bibitem{structure}
{\it R}Te$_{2}$ crystallises in space group number 129 (P4/nmm). The internal
parameters are {\it R}(1):~0.25~0.25~0.2744; Te(1):~0.75~0.25~0.5; Te(2):~
0.25~0.25~0.8735 (Ref. \cite{dimasi:96}). {\it R}Te$_{3}$ crystallises in space
group number 63 (Cmcm). The internal parameters are {\it R}(1):~0~0.1694~0.25; Te(1):~
0~0.92925~0.25; Te(2):~0~0.5706~0.25; Te(3):~0~0.2953~0.25 (Ref. \cite{norling:66}).

\bibitem{kikuchi:97}
A.~Kikuchi, J. Phys. Soc. Jap. {\bf 67}, 1308 (1997)


\bibitem{west:95}
R.~N. West,  in {\em Proceedings of the International School of Physics
  $<<$ Enrico Fermi $>>$ --- Positron Spectroscopy of Solids}, edited by A.
    Dupasquier and A.~P.~Mills jr. (IOS Press, Amsterdam, 1995).

\bibitem{dugdale:97}
See, for example
S.~B.~Dugdale, M.~A.~Alam, I.~Wilkinson, R.~J.~Hughes, I.~R.~Fisher, P.~C.~Canfield,
T.~Jarlborg and G.~Santi, Phys. Rev. Lett. {\bf 83}, 4824 (1999)~;
Zs.~Major, S.~B.~Dugdale, R.~J.~Watts, G.~Santi, M.~A.~Alam, S.~M.~Hayden,
J.~A.~Duffy, J.~W.~Taylor, T.~Jarlborg, E.~Bruno, D.~Benea and H.~Ebert,
Phys. Rev. Lett. {\bf 92}, 107003 (2004).

\bibitem{lock:73}
D.~G. Lock, V.~H.~C. Crisp and R.~N. West, J. Phys. F {\bf 3},  561 (1973).

\bibitem{dugdale:94}
S.~B. Dugdale, M.~A. Alam, H.~M. Fretwell, M. Biasini and D. Wilson, 
J. Phys.:Condens. Matt. {\bf 6}, L435 (1994).

\end{thebibliography}
\end{document}